\documentclass[]{aa}  

\usepackage{natbib}
\bibpunct{(}{)}{;}{a}{}{,}

\usepackage{graphicx}
\usepackage{revsymb}	
\usepackage{amsmath}
\usepackage{txfonts}
\usepackage{amssymb}
\usepackage{geometry} 
\usepackage[parfill]{parskip} 
\usepackage{slantsc}
\usepackage{array}
\usepackage{amsfonts}
	
\usepackage{epstopdf}	
\usepackage{epsfig}	

\usepackage{url}

\usepackage[colorlinks=true,linkcolor=black, urlcolor=black, citecolor=black]{hyperref}

\usepackage{color}
\usepackage{xcolor}
\colorlet{rouge}{red!70!darkgray}

\usepackage{xfrac}

\usepackage{sidecap}
\usepackage[font=small, labelfont=bf]{caption}
\usepackage{subcaption}

\usepackage{threeparttable}

\begin{document}
   \title{Probing stellar cores from inversions of frequency separation ratios}
\author{J. B\'{e}trisey\inst{1} \and G. Buldgen\inst{1}}
\institute{$^1$Observatoire de Genève, Université de Genève, Chemin Pegasi 51, 1290 Versoix, Suisse.}

\date{March, 2022}

\abstract{With the rapid development of asteroseismology thanks to space-based photometry missions such as CoRoT, \emph{Kepler}, TESS, and in the future, PLATO, and the use of inversion techniques, quasi-model-independent constraints on the stellar properties can be extracted from a given stellar oscillation spectrum. In this context, inversions based on frequency separation ratios, that are less sensitive to surface effects, appear as a promising technique to constrain the properties of stellar convective cores.}
{We developed an inversion based on frequency separation ratios with the goal of damping the surface effects of the oscillation frequencies. Using this new inversion, we defined a new indicator to constrain the boundary mixing properties of convective cores in solar-like oscillators.}
{We verified our inversion technique by conducting tests in a controlled environment, where the stellar mass and radius are known exactly, and conducted an extensive hare and hounds exercise.}
{The inversion is not affected by surface effects. With the construction of an extensive set of models, favoured and forbidden regions can be highlighted in the parameter space. If the ratios are well fitted, the inversion is unsurprisingly not providing additional information.}
{The indicator coupled with the inversion based on frequency separation ratios seems promising at probing the properties of convective cores, especially for F-type stars exhibiting solar-like oscillations.}


\maketitle

\section{Introduction}
In the last two decades, the space-based photometry missions CoRoT \citep{Baglin2009}, \emph{Kepler} \citep{Borucki2010}, and TESS \citep{Ricker2015} were launched, enabling a rapid development of asteroseismology. New progresses are also expected from the future PLATO mission \citep{Rauer2014}. The quality of the data made possible the use of so-called seismic inversion techniques, until then applied with tremendous success in helioseismology \citep[see][for recent reviews]{Buldgen2019e,JCD2021}. The situation in asteroseismology is however very different from that of the Sun, as geometric cancellation forbids the detection of modes with angular degrees higher than 3 for solar-like oscillators. This limits the available datasets to about 50 frequencies at best, whereas thousands of modes are observed on the Sun. As a consequence, the philosophy of inversions was reoriented. Indeed, the classical inversions for the Sun are local inversions that can scan through the entire structure, but they require a large set of frequencies to be efficient, which can only be achieved for the best asteroseismic targets \citep[e.g. 16CygA and 16CygB,][]{Bellinger2017,Bellinger2019d,Bellinger2021,Buldgen2022b}, and with the restriction that information can only be extracted for the deep layers. Therefore, inversions based on so-called indicators were developed \citep{Reese2012,Buldgen2015a,Buldgen2015b,Buldgen2018}, and applied to various cases \citep[e.g.][]{Buldgen2016b,Buldgen2016c,Buldgen2017c,Buldgen2019b,Buldgen2019f,Betrisey2022}. Here, the idea is to concentrate all the information of the frequency spectrum to conduct a global inversion that constrains a global quantity, the indicator (e.g. the mean density).

In this work, we are interested in the entropy profile in the central adiabatic convective regions to put some constraints on the properties of the convective core. With the current inversions based on individual frequencies, a large number of them are required, especially at various harmonic degrees. In addition, a very accurate characterisation of the mass and radius is also needed \citep{Buldgen2017a}. Entropy inversions were conducted for the Sun \citep{Buldgen2017d}, and for the 16Cyg binary system \citep{Buldgen2022b} using the indicators defined in \citet{Buldgen2018}. However, even though the 16Cyg system is one of the best \textit{Kepler} asteroseismic target, \citet{Buldgen2022b} found that no additional information about the central entropy could be extracted for 16CygB. In addition, inversions based on individual frequencies are sensitive to the surface regions, which is a limiting factor as the near-surface effects are poorly modelled. In the case of 16CygA, the surface effects and the quality of the reference models prevented \citet{Buldgen2022b} to extract new constraints about the central entropy. Therefore, we chose to develop a new indicator based on frequency separation ratios instead of individual frequencies, aiming to carry out inversions of solar-like stars presenting convective cores. Indeed, frequency separation ratios are defined to damp surface effects \citep{Roxburgh&Vorontsov2003} and are more efficient at probing central stellar regions \citep{Oti2005}. Furthermore, \citet{Deheuvels2016} and \citet{Farnir2019} have shown that frequency separation ratios could be used to constrain the extent of convective cores in solar-like oscillators.

In Sec. \ref{sec_theoretical_considerations} we recall the state of the art for indicator-based inversions using individual frequencies, and introduce a new inversion based on frequency separation ratios. In Sec. \ref{sec_new_indicator_probing_cores} we define a new indicator probing central stellar regions and test it in a controlled environment. In Sec. \ref{sec_H&H} we conduct a hare and hounds exercise to simulate how the indicator would behave with a real observed target. Finally, in Sec. \ref{sec_conclusions} we draw the conclusions of our study.

\section{Theoretical considerations}
\label{sec_theoretical_considerations}
In the inversion community, the following terminology is in general applied. The inversion aims at retrieving the properties of an \textit{observed/target} model, which can either be a real observation, or a synthetic target. To this end, the inversion takes as input a \textit{reference} model, which is the result of a modelling procedure, and outputs a small correction that is added to a quantity of interest computed from the reference model. For simplicity in this manuscript, an \textit{inverted} quantity is defined as the quantity of interest that includes the correction from the inversion.

\subsection{State of the art for individual frequencies}
The structure inversion equation is based on a perturbative analysis of the stellar oscillations at linear order. This approach is motivated by the work of \citet{Lynden-Bell&Ostriker1967} and their predecessors \citep[see e.g.][]{Chandrasekhar1964,Chandrasekhar&Lebovitz1964,Clement1964}, who showed that the equation of motion fulfils a variational principle. In the case of individual frequencies, the frequency perturbation is directly related to the structural perturbation \citep{Dziembowski1990}:
\begin{equation}
\frac{\delta\nu^{n,l}}{\nu^{n,l}} = \int_{0}^{R} K_{a,b}^{n,l}\frac{\delta a}{a}dr + \int_{0}^{R} K_{b,a}^{n,l}\frac{\delta b}{b}dr + \mathcal{O}(\delta^2),
\label{eq_inversionGT}
\end{equation}
with $\nu$ the oscillation frequency, $n$ the radial order, $l$ the harmonic degree, $a$ and $b$ two structural variables, $K_{a,b}^{n,l}$ and $K_{b,a}^{n,l}$ the structural kernels and using the definition
\begin{equation}
\frac{\delta x}{x} = \frac{x_{\mathrm{obs}}-x_{\mathrm{ref}}}{x_{\mathrm{ref}}},
\end{equation}
where the index $ref$ stands for reference, and $obs$ stands for observed. Historically, Eq. \eqref{eq_inversionGT} was derived for the sound speed and density structural pair $(c^2,\rho)$, but the equation can be generalized to any combination of physical quantities appearing in the adiabatic oscillation equations \citep{Elliott1996,Kosovichev2011,Buldgen2017a}.

Given the frequency differences, the idea is then to solve Eq. \eqref{eq_inversionGT} to have access to the structure differences, and compute a global integrated quantity, the so-called indicator $t$ (e.g. the mean density), which concentrates all the information of the frequency spectrum. The general form of the indicator is 
\begin{equation}
t = \int_0^R f(r)g(a)h(b)dr,
\end{equation}
where $f$ is a weight function that depends on the radius, and $g$ and $h$ are two functions of the structural variables. Expect in rarer cases \citep[see e.g.][]{Buldgen2015a}, $h(b)$ is set to one, and the indicator is only a function of one of the structural variables. In practice, the options for $g$ are limited, because the target function $\mathcal{T}_t$ in Eq. \eqref{eq_SOLA} must be reproducible with the number of modes available in asteroseismology. Therefore, simple functions such as $g(a)=a$, $g(a)=1/a$, etc. are good candidates. A more detailed explanation of the determination of $f$ and $g$ is provided in Sec. \ref{sec_new_indicator_probing_cores}, where we define the indicator used in this work.

This approach has the advantage of being quasi-model independent, because it does not rely on the physics assumed to generate the reference model that serves as starting point for the inversion. In addition, the inversion is also quasi-independent from the starting point in the sense that if we start from another point in the parameter space, the inversion will still correct towards the exact value.

The inversion can be performed with the Subtractive Optimally Localized Averages (SOLA) method \citep{Pijpers&Thompson1994}. This minimization technique is an adaptation of the OLA approach of \citet{Backus&Gilbert1968,Backus&Gilbert1970}, and allows to conduct the inversion in a numerically more efficient way. In practice, for a given indicator $t$, the following cost function is minimized:
\begin{align}
\mathcal{J}_t(c_i) &= \int_0^1 \big(\mathcal{K}_{\mathrm{avg}} - \mathcal{T}_t\big)^2 dx 
                                        + \beta\int_0^1 \mathcal{K}_{\mathrm{cross}}^2dx + \lambda\left[k-\sum_i c_i\right] \nonumber\\
                                        &\quad+\tan\theta \frac{\sum_i (c_i\sigma_i)^2}{\langle\sigma^2\rangle}
                                        + \mathcal{F}_{\mathrm{Surf}}(\nu),
\label{eq_SOLA}
\end{align}
where the averaging and cross-term kernels are related to the structural kernels,
\begin{align}
\mathcal{K}_{\mathrm{avg}} &= \sum_i c_i K_{a,b}^{i}, \\
\mathcal{K}_{\mathrm{cross}} &= \sum_i c_i K_{b,a}^{i}.
\end{align}

The normalization $k$ depends on the properties of the indicator. The balance between the amplitudes of the different terms during the fitting is adjusted with trade-off parameters, $\beta$ and $\theta$. Here, the idea is to provide a good fit of the target function $\mathcal{T}_t$ while reducing the contribution of the cross-term and of the observational errors on the individual frequencies. We introduce a short notation for the identification pair of a frequency, $i\equiv (n,l)$. The $1\sigma$-uncertainty of the relative frequency differences is denoted by $\sigma_i$, and $\langle\sigma^2\rangle = \sum_i^N\sigma_i^2$ is defined, with $N$ the number of observed frequencies. The variable $\lambda$ is a Lagrange multiplier, and $c_i$ are the inversion coefficients.

The function $\mathcal{F}_{\mathrm{Surf}}(\nu)$ is an empirical term to describe the surface effects. Besides introducing additional free parameters in the minimization at the expense of the fit of the target function, the treatment of the surface effects constitutes a limitation of the inversion. Indeed, structural kernels based on individual frequencies have a large amplitude in the surface regions, as shown in the upper panel of Fig. \ref{fig_kernels}, and are therefore sensitive to the treatment of these layers. In practice, this limitation can be translated into a systematic uncertainty \citep[see e.g.][]{Betrisey2022}, but it can become significant enough to prevent the inversion of accurately correcting.

\subsection{Adaptation for frequency separation ratios}
\label{sec_adaptation_frequency_ratios}
\citet{Roxburgh&Vorontsov2003} introduced frequency separation ratios to damp surface effects. They are defined as the ratio of the small frequency separations divided by the large frequency separations. Because the modes with higher harmonic degree than $l=2$ are difficult to observe in asteroseismology, we will only consider the following ratios:
\begin{align}
r_{01}(n) &= \frac{d_{01}(n)}{\Delta_1(n)}, \\
r_{10}(n) &= \frac{d_{10}(n)}{\Delta_0(n+1)}, \\
r_{02}(n) &= \frac{d_{02}(n)}{\Delta_1(n)},
\end{align}
where $\Delta_l(n)$ are the large separations, and $d_{xy}(n)$ are the small separations:
\begin{align}
\Delta_l(n) &= \nu_{n,l}-\nu_{n-1,l}, \\
d_{01}(n) &=  \frac{1}{8}\left(\nu_{n-1,0}-4\nu_{n-1,1}+6\nu_{n,0}-4\nu_{n,1}+\nu_{n+1,0}\right), \\
d_{10}(n) &=  -\frac{1}{8}\left(\nu_{n-1,1}-4\nu_{n,0}+6\nu_{n,1}-4\nu_{n+1,0}+\nu_{n+1,1}\right), \\
d_{02}(n) &= \nu_{n,0}-\nu_{n-1,2}.
\end{align}

The first version of a kernel of a ratio was presented in \citet{Oti2005}, with the introduction of the $r_{02}$ kernels, and modulo a different scaling than in this work. We present here the derivation of the $r_{01}$ kernels. The derivations for the $r_{10}$ and $r_{02}$ kernels can be found in Appendix \ref{sec_appendix_kernel_r10_r02}. As for the individual frequencies, the relative differences of the ratios can be related to the structure differences
\begin{align}
\frac{\delta r_{01}}{r_{01}}(n) &= \frac{\delta d_{01}(n)}{d_{01}(n)} - \frac{\delta\Delta_1(n)}{\Delta_1(n)}, \\
                                                        &= \int_0^R\left(K_{a,b}^{r_{01}}(n)\frac{\delta a}{a} + K_{b,a}^{r_{01}}(n)\frac{\delta b}{b}\right)dr
                                                               + \mathcal{O}(\delta^2).
\end{align}
Because a frequency ratio is a linear combination of individual frequencies, its kernel is a linear combination of the structural kernels of the frequencies used in the definition of the ratio. For example, the $r_{01}$ kernels are the result of the following combination:
\begin{small}
\begin{align}
K_{a,b}^{r_{01}}(n) = &\frac{\nu_{n-1,0}K_{a,b}^{n-1,0}-4\nu_{n-1,1}K_{a,b}^{n-1,1}+6\nu_{n,0}K_{a,b}^{n,0}
                                                  -4\nu_{n,1}K_{a,b}^{n,1}+\nu_{n+1,0}K_{a,b}^{n+1,0}}{\nu_{n-1,0}-4\nu_{n,1}+6\nu_{n,0}
                                                  -4\nu_{n,1}+\nu_{n+1,0}} \nonumber\\
                                        &   - \frac{\nu_{n,1}K_{a,b}^{n,1}-\nu_{n-1,1}K_{a,b}^{n-1,1}}{\nu_{n,1}-\nu_{n-1,1}}.
\end{align}
\end{small}

The kernels $K_{b,a}^{r_{01}}(n)$ are computed by switching $a$ and $b$ in the previous expression.

\begin{figure}[h!]
\centering
\includegraphics[scale=0.4]{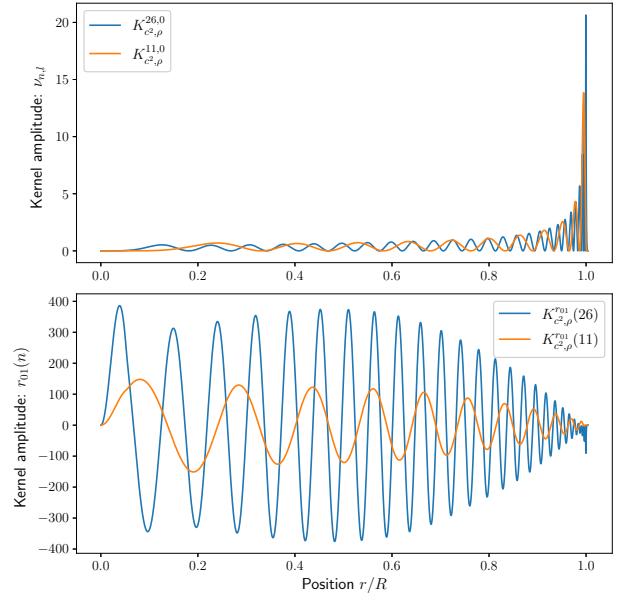} 
\caption{Frequency kernels versus ratio kernels of the model \textit{1.3MoAGSS} from Sec. \ref{sec_definition_indicator}. For both panels, we show the kernel amplitude as a function of the position, and for the structural variable pair $(c^2,\rho)$. \textit{Upper panel:} Low order ($l=0$, $n=11$) and high order ($l=0$, $n=26$) kernels for the individual frequencies, respectively in orange and blue. \textit{Lower panel:} Low order ($n=11$) and high order ($n=26$) kernels for the frequency separation ratios, respectively in orange and blue.}
\label{fig_kernels}
\end{figure}

Fig. \ref{fig_kernels} shows a comparison of the kernels for the individual frequencies and for the frequency separation ratios for a 1.3M$_\odot$ star (model \textit{1.3MoAGSS} from Sec. \ref{sec_definition_indicator}). The frequency kernels have a large amplitude at the surface of the star. They are therefore well suited to probe this region, but are sensitive to surface effects. On the contrary, ratio kernels have an amplitude that is damped at the surface, suppressing the surface effects. In addition, they have a larger amplitude in the central stellar regions, allowing to probe these regions more efficiently. We chose to display the kernels of the structural pair $(c^2,\rho)$ because it nicely illustrates this behaviour, but in practice, another structural pair should be used. Indeed, the form of these kernels is not adapted with our indicator, and the contribution of the cross-term is important for this structural pair. For inversions based on frequency separation ratios, structural pairs such as $(S_{5/3},Y)$, or $(S_{5/3},\Gamma_1)$ are more suited. The variable $S_{5/3}=P/\rho^{5/3}$ is a proxy of the entropy, $P$ is the pressure, $\rho$ is the density, $Y$ is the helium mass fraction, and $\Gamma_1=\left(\frac{\partial \ln P}{\partial \ln\rho}\right)_{\mathrm{ad}}$ is the first adiabatic exponent.

There are two main differences for the SOLA cost function. We cannot assume that the ratio uncertainties are uncorrelated because a ratio is by definition a combination of frequencies, and the surface effects can be neglected by construction (as we will show later).
\begin{align}
\mathcal{J}_t(c_i) &= \int_0^1 \big(\mathcal{K}_{\mathrm{avg}} - \mathcal{T}_t\big)^2 dx 
                                   + \beta\int_0^1 \mathcal{K}_{\mathrm{cross}}^2dx \nonumber\\
                                    &\quad+\tan\theta\sum_{p=1}^{N_r}\sum_{q=1}^{N_r}\frac{c_p c_q \vec{\Sigma}_{pq}}{<\vec{\Sigma}^2>}.
\label{eq_SOLA_ratios}
\end{align}

As for the individual frequencies, $\beta$ and $\theta$ are trade-off parameters to balance the amplitudes of the different terms during the fitting. Again, the idea is to provide a good fit of the target function $\mathcal{T}_t$ while reducing the contribution of the cross-term and of the observational errors. The variable $\lambda$ is a Lagrange multiplier, and $c_p$ are the inversion coefficients. The normalisation $<\vec{\Sigma}^2> = \frac{1}{N_r^2}||\vec{\Sigma}||_\mathrm{F}^2$, where $||\vec{\Sigma}||_\mathrm{F}^2$ is the square of the Frobenius norm of the covariance matrix of the ratios, and $N_r$ is the number of observed ratios. The entries of the covariance matrix of the ratios are given by Eq. \eqref{eq_entries_cov_matrix_ratios} from Appendix \ref{sec_appendix_correlation_matrix_ratios}: $\vec{\Sigma}_{pq}=\mathrm{Cov}\left((\frac{\delta r}{r})_p,(\frac{\delta r}{r})_q\right)$.

\section{A new indicator probing stellar cores}
\label{sec_new_indicator_probing_cores}
In this section, we aim at defining an indicator that is sensitive to the properties of convective cores. For the first variable, the entropy proxy $S_{5/3}$ appears as a natural choice. Indeed, this proxy is motived by the Sackur-Tetrode equation for the entropy in the case of an ideal gas,
\begin{align}
S = \frac{3k_B}{2}\left[\mu m_u \ln\left(\frac{P}{\rho^{5/3}} \right) + g(\mu) \right],
\end{align}
where $k_B$ is the Boltzmann constant, $\mu$ is the mean molecular weight, $m_u$ is the atomic mass unit, and $g(\mu)$ is a function that depends only on the mean molecular weight. They both form a plateau in adiabatic convective regions, and the height and extension of this plateau is related to the boundary mixing properties, such as the temperature and mean molecular weight gradients. Because these boundary regions are generally poorly modelled, an indicator sensitive to their properties can provide new constraints on the physical processes acting in these regions. 

For the second structural variable, we chose the helium mass fraction. In comparison to $\Gamma_1$, the amplitude of the cross-term is significantly smaller, which naturally reduces the cross-term contribution in the SOLA cost function. Even if the structural pair including $\Gamma_1$ is less suited, an inversion based on these variables is still feasible, but with a lessened stability. Indeed, we can define error measures to quantify the sources of error in an inversion \citep{Buldgen2015b}. There are the error on the averaging and cross-term kernels, which quantifies how well the associated target functions are reproduced. The last source of error is the residual error that quantifies the unknown remaining error after taking into account the two previous error measures. In an ideal case, the averaging kernel perfectly reproduces the target function, the cross-term is zero, and the inversion can therefore retrieve the exact value. In practice, the cross-term error is negligible, and compensations occur when the averaging and residuals errors are of the same magnitude but with opposite sign. In that case, the inversion is unstable because a significant unknown source of error affects the inversion result.

In the following, we will define a new indicator probing central stellar regions, and test it in a controlled environment where the stellar mass and radius between the reference and target models are identical. In these conditions, the mean density is well controlled, which strengthens the stability of the inversion.

\subsection{Definition and calibration}
\label{sec_definition_indicator}
We define our indicator as follows:
\begin{align}
S_{core}^{r_{01}} = \int_0^R \frac{-f(r)}{S_{5/3}(r)}dr,
\end{align}
where $f(r)$ is a weight function. The minus sign is purely cosmetic to ensure that the indicator is positive. Because only a small number of modes are available in asteroseismology, the target function must be constructed so that it can be fitted by the structural kernels. Therefore, the weight function is chosen such that the resulting target function is close to the natural form of the kernels.

We compute the eulerian linear perturbation of the indicator:
\begin{align}
\frac{\delta S_{core}^{r_{01}}}{S_{core}^{r_{01}}} 
&= \int_0^R \frac{f(r)}{S_{core}^{r_{01}}\cdot S_{5/3}(r)}\frac{\delta S_{5/3}(r)}{S_{5/3}(r)} dr,\\
& = \int_0^R\mathcal{T}_{S_{core}^{r_{01}}}(r) \frac{\delta S_{5/3}(r)}{S_{5/3}(r)} dr.
\end{align}
Hence, the target function is:
\begin{align}
\mathcal{T}_{S_{core}^{r_{01}}}(r) = \frac{f(r)}{S_{core}^{r_{01}}\cdot S_{5/3}(r)}.
\end{align}

\begin{figure}[h!]
\centering
\includegraphics[scale=0.55]{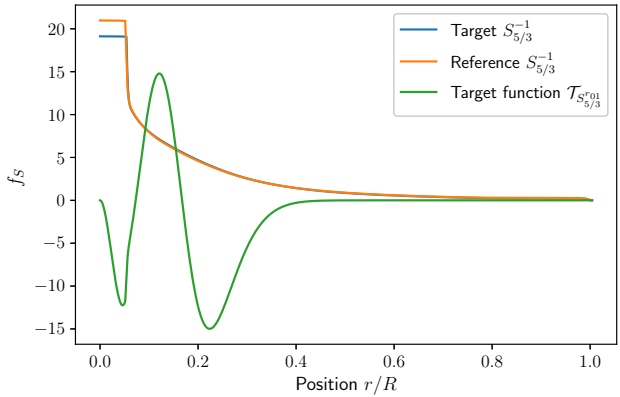}
\caption{Target function and entropy profiles for the calibration model. The observed and modelled entropy profiles are in blue and orange, respectively. The target function is in green.}
\label{fig_profile_vs_target_function}
\end{figure}

The determination of the weight function is the most difficult part when developing a new indicator. Indeed, there is no straightforward procedure to determine it, except trial and error, and one has to find a good balance between the fine-tuning of the weight function and the trade-off parameters. In practice, to find the weight function, we start with a guess for $f(r)$ and iterate until we find a target function that is reproducible with the available kernels. There is no receipt for the initial guess, but we started with a $r^2$ term because the kernels follow approximately this behaviour in the very central regions, an exponential decay to suppress the signal at the surface, and two Gaussian functions. After a few iterations, we found that including a third Gaussian function to match the number of "lobes" of the target function, illustrated in Fig. \ref{fig_profile_vs_target_function}, as well as the hyperbolic tangent to suppress more efficiently the signal until the center of the star, produces better results. Because we use $Y$ as structural variable, the cross-term is naturally damped and the value of $\beta$ has very little impact on the inversion. To find an adequate balance, we proceeded as follows. We calibrated the weight function and the trade-off parameters for one model, the model \textit{1.3MoAGSS} trying to reproduce the $\rm1.3M_\odot$ synthetic target of the next section. Then, we verified the indicator in a controlled environment (Sec. \ref{sec_behaviour_controlled_environment}), and with an hare and hounds exercise (Sec. \ref{sec_H&H}). We found that adopting $\beta=10^{-4}$, $\theta=10^{-4}$, and
\begin{align}
f(r)= r^2 e^{-13r} \Big(f_1(r)+f_2(r)+f_3(r)\Big)\tanh\left(3(1-r)^7\right),
\end{align}
where
\begin{align}
f_1(r) &= -\exp\left(-150(r-0.02)^2\right), \\
f_2(r) &= 3\exp\left(-150(r-0.15)^2\right), \\
f_3(r) &= -6\exp\left(-50(r-0.29)^2\right),
\end{align}
is a good compromise in most of the cases.

In Fig. \ref{fig_profile_vs_target_function}, we show the target function and the entropy profiles for the calibration model to illustrate where the information is localised and to which stellar regions the target function is the most sensitive. The structural differences are mainly located in the central region, up to 1/10$^{th}$ of the stellar radius. Therefore, most of the contribution in the inversion will originate from this region. The first inner peak and the inner part of the second peak of the target function are the most important because they can "see" where the structural differences are localised. Also, it is not necessary to reproduce efficiently the last peak of the target function because it is sensitive to a region where the structural differences are almost non-existent. In addition, we recall that the kernels are basis vectors that are combined to reproduce the target function. From a linear algebra point of view, some of them will generate the global shape of the target function, while some others will reproduce the details. In our inversion, the global shape is generated by the low-order kernels and they have the highest coefficients. The high-order kernels retrieve some of the details and have smaller coefficients. In this sense, low-order kernels are thus an asset for the inversion. We note that extreme orders ($n<5$ or $n>30$) showed a numerically unstable behaviour. This might be a numerical effect that could be fixed or the sign of an intrinsic non-linear behaviour. From a theoretical point of view, further investigations beyond the scope of this article are required, but modes with $n<5$ are currently unrealistic to measure, and high-order modes ($n>30$), in addition of also being very difficult to observe, will be very sensitive to surface effects.

\begin{figure}[h!]
\centering
\includegraphics[scale=0.45]{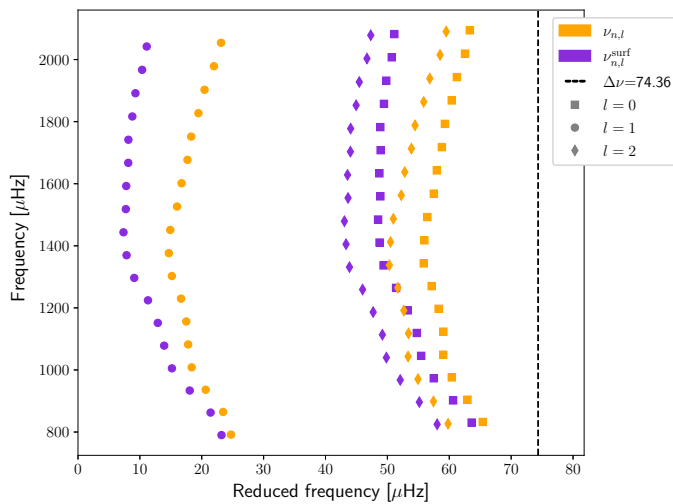} 
\caption{Impact of the surface effects on the frequencies. The frequencies before (orange) and after (purple) the addition of the surface effects are shown for different harmonic degrees $l$.}
\label{fig_freq_sonoi}
\end{figure}

For the calibration model, we tested that the inversion is indeed not affected by surface effects. To this end, we added surface effects on the frequencies of the synthetic observed model using the \citet{Sonoi2015} prescription, which generates strong surface corrections, especially at high order, as shown in Fig. \ref{fig_freq_sonoi}. Even though the frequencies are significantly affected, the impact on the ratios is very small, and the relative ratio differences are modified on average by about 0.08\%, and much less for low order modes that carry the inversion. Because this effect is much smaller than any optimistic observational uncertainty, especially at low order, the impact of the surface effects on the inversion is completely negligible.

\subsection{Behaviour in a controlled environment}
\label{sec_behaviour_controlled_environment}
For the preliminary tests in a controlled environment, we generated synthetic "observed" targets in the main sequence with masses of $\rm 1.1M_\odot$, $\rm 1.2M_\odot$, $\rm 1.3M_\odot$, $\rm 1.4M_\odot$, and $\rm 1.5M_\odot$, and modified one physical ingredient at a time. For the observed targets, we used the OPAL opacities \citep{Iglesias1996}, supplemented by the \citet{Ferguson2005} opacities at low temperature and the electron conductivity by \citet{Potekhin1999}. We used the GN93 abundances \citep{Grevesse&Noels1993}, the FreeEOS equation of state \citep{Irwin2012}, and microscopic diffusion was described using the formalism of \citet{Thoul1994}, but with the screening coefficients of \citet{Paquette1986}, and taking partial ionization into account. The nuclear reaction rates are from \citet{Adelberger2011}. We fixed the mixing-length parameter $\alpha_{\mathrm{MLT}}$ at a solar calibrated value of 2.05, and following the implementation of \citet{Cox&Giuli1968}. For the atmosphere modelling, we used the $T(\tau)$ relation from \citet{Eddington1959}. For all the models in this work, we used the Liège Evolution Code \citep[CLES,][]{Scuflaire2008b}, and the frequencies were computed with the adiabatic Liège Oscillation Code \citep[LOSC,][]{Scuflaire2008a}.

For the reference models, we changed the abundances for the AGSS09 abundances \citep{Asplund2009}, the opacities for the OPLIB opacities \citep{Colgan2016}, and the atmosphere by using the $T(\tau)$ relation described by Model-C in \citet{Vernazza1981} (hereafter VAL-C). We also introduced overshooting and undershooting, respectively with $\rm\alpha_{ov}=0.1$ and $\rm\alpha_{under}=0.3$, and assuming an adiabatic stratification in both cases. We selected the reference models in the evolutionary sequence by matching the radius with the observation. Concerning the uncertainty of the relative ratios differences, we adopted an error of 20\% (hereafter equal-weighted uncertainties) that corresponds to the expectation for high quality data.

\begin{figure}[h!]
\centering
\includegraphics[scale=0.4]{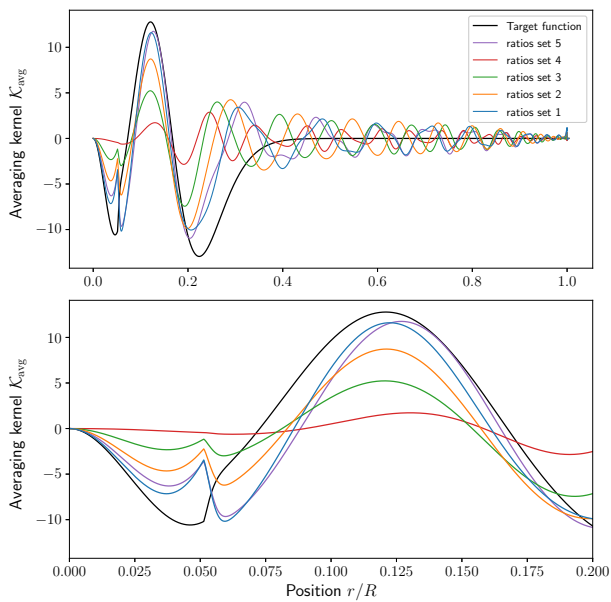}
\caption{Averaging kernel considering several sets of modes. \textit{Upper panel:} Averaging kernels over the full radius domain. \textit{Lower panel:} Zoom on the region that contributes to the inversion.}
\label{fig_target_function_nb_modes}
\end{figure}

For the inversions, we considered several sets of modes, summarised in Table \ref{tab_mode_set}. As shown in Fig. \ref{fig_target_function_nb_modes}, with the addition of low order modes, the averaging kernel better reproduces the target function, especially in the central regions. Moreover, high order modes only have a small contribution. Indeed, the target function of sets 1 and 5 is almost identical. For the sets 3 and 4, the target function is poorly reproduced by the averaging kernel, and it is meaningless to carry out an inversion in such unstable conditions. In practice, we would recommend to use mode sets similar to sets 1, 2, or 5, and stress out that such mode sets can be observed (and were for some \emph{Kepler} targets for example).

\begin{table}[h!]
\centering
\caption{Mode sets for the inversions in a controlled environment.}
\begin{tabular}{lc}
\hline 
 & radial order $n$ \\ 
\hline \hline 
set 1 & 11-26 \\ 
set 2 & 13-26 \\ 
set 3 & 15-26 \\  
set 4 & 18-26 \\ 
set 5 & 11-23 \\ 
\hline
\end{tabular} 
\par
 \vspace{1ex}
{\small\raggedright\textbf{Notes.} For the models with $\rm 1.4M_	\odot$ and $\rm 1.5M_	\odot$, $n_{max}=27$ (instead of 26). \par}
\label{tab_mode_set}
\end{table}

\begin{figure*}[h!]
\centering
\includegraphics[scale=0.85]{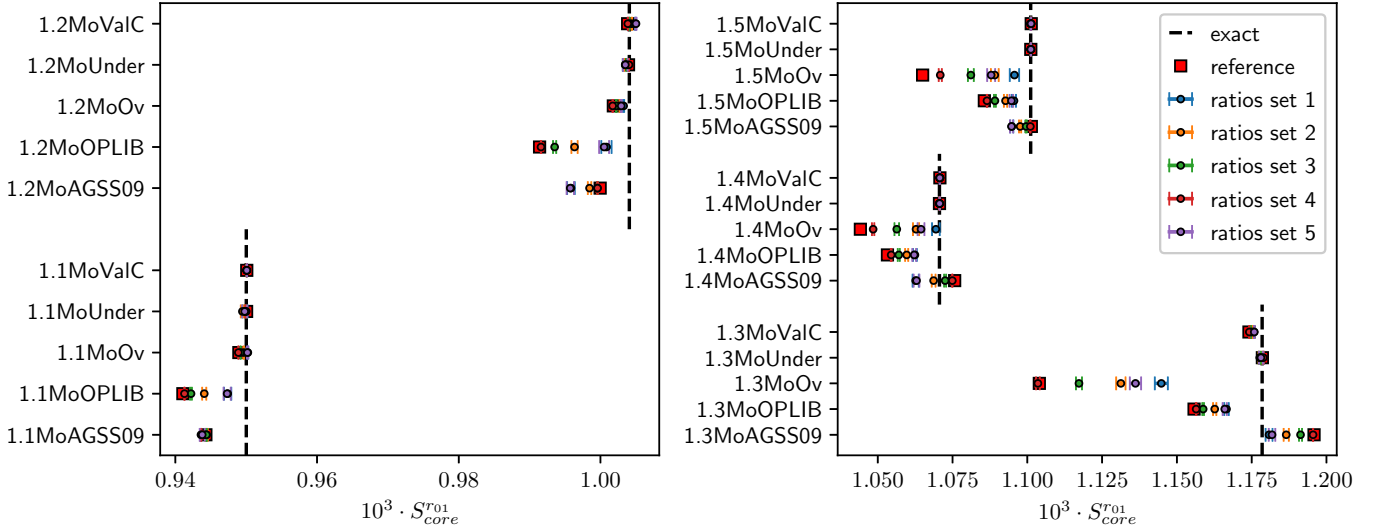}
\caption{Behaviour of the indicator in a controlled environment. \textit{Left panel:} Results for the lower masses ($\rm 1.1M_\odot$ to $\rm 1.2M_\odot$). \textit{Right panel:} Results for the higher masses ($\rm 1.3M_\odot$ to $\rm 1.5M_\odot$).} 
\label{fig_phase1_results}
\end{figure*}

The inversion results are shown in Fig. \ref{fig_phase1_results}, for lower mass stars ($\rm 1.1M_\odot$ to $\rm 1.2M_\odot$), and higher mass stars ($\rm 1.3M_\odot$ to $\rm 1.5M_\odot$). The atmosphere model and undershooting affect surface regions, and have no impact on the central entropy profile. The indicator is therefore insensitive to these quantities. It shows promising results for the abundances, and especially for the opacities and the overshoot. The latter indeed induce more significant structural changes, especially for the higher masses, which can be captured by the inversion. For the models with the AGSS09 abundances and masses of $\rm 1.2M_\odot$, $\rm 1.4M_\odot$, and $\rm 1.5M_\odot$, the inversion was less or not successful because the relative ratio differences were too small. An inversion has a resolution limit linked with the averaging error and how well the target function is fitted. If the reference model is too close to the observed model, the error on the averaging kernel dominates the inversion, and prevent a meaningful correction. Hence, if the ratios are well fitted, it is not appropriate to carry out an inversion.

\section{Hare and hounds}
\label{sec_H&H}

\begin{table*}
\centering
\caption{Characteristics of the hares.}
\label{tab_hares_prop}
\begin{tabular}{lcccccc}
\hline 
 & Target 1 & Target 2 & Target 3 & Target 4 & Target 5 & Target 6 \\ 
\hline \hline
Mass ($M_\odot$) & 1.10 & 1.26 & 1.42 & 1.48 & 1.35 & 1.23 \\ 
Age (Gyr) & 3.70 & 3.55 & 1.67 & 2.09 & 2.09 & 3.16 \\ 
Radius ($R_\odot$) & 1.165 & 1.515 & 1.734 & 2.274 & 1.816 & 1.444 \\ 
$T_{eff}\ (K)$ & 5978 & 5951 & 6604 & 6374 & 6629 & 6212 \\ 
$Z_0$ & 0.020 & 0.028 & 0.023 & 0.015 & 0.0135 & 0.020 \\ 
$X_0$ & 0.71 & 0.68 & 0.69 & 0.70 & 0.685 & 0.71 \\ 
Abundances & GN93 & AGSS09 & GN93 & AAG21 & AGSS09 & GN93 \\ 
Opacities & OPAL & OPAL & OPLIB & OPAL & OPAL & OPAL \\ 
EOS & FreeEOS & FreeEOS & FreeEOS & FreeEOS & FreeEOS & FreeEOS \\
$\rm\alpha_{Conv}$ & 1.90 & 1.75 & 2.15 & 2.05 & 0.85 & 1.90 \\ 
$\rm\alpha_{Ov}$ & 0.05 & 0.15 & 0.10 & 0.15 & 0.13 & 0.10 \\ 
$\rm\alpha_{Under}$ & 0.10 & 0.15 & 0.10 & 0.10 & 0.10 & 0.10 \\ 
Convection & MLT & MLT & MLT & MLT & FST & MLT \\ 
Microscopic diffusion & YES & YES & YES & YES - only X, Y & YES & YES \\ 
Turbulent diffusion & YES & NO & YES & NO & YES & YES \\
$D_{T}$ & 7500 & 0 & 2500 & 0 & 50 & 7500 \\ 
$\beta$ & 3 & 0 & 2 & 0 & 2 & 3 \\ 
Atmosphere & VAL-C & VAL-C & VAL-C & VAL-C & VAL-C & VAL-C \\
Radial order $n$ & 13-25 & 13-25 & 11-23 & 11-25 & 12-23 & 10-22 \\
\hline 
\end{tabular} 
\end{table*}
\subsection{Presentation of the hares}
In this hare and hounds exercise, we aimed at computing models with different physical ingredients, unknown to the hounds, to reproduce as realistically as possible the presence of additional physical processes in observed targets unaccounted for in the seismic modelling procedure. To avoid biases, the hares were generated independently, and only the frequencies, as well as the spectroscopic constraints (effective temperature, surface metallicity, and luminosity), were provided to the modeller of the hounds. After all the inversions were carried out, the modeller got access to the stellar properties and internal profiles of the hares to draw the conclusions of this hare and hounds exercise.

The properties of the hares are presented in Table \ref{tab_hares_prop}. We varied the opacity tables, using either the OPAL or the OPLIB opacities, the reference solar abundances using either the AGSS09 or the GN93 abundances, the formalism for convection, using either the classical mixing-length theory or the full-spectrum of turbulence theory \citep{Canuto1996}. The effects of microscopic and turbulent diffusion were considered, with one hare only taking into account diffusion of helium and hydrogen, as to simulate the effects of radiative accelerations on metals. Other ingredients were kept unchanged such as the equation of state, for which we used the FreeEOS equation of state and the atmosphere model, for which we used a VAL-C atmosphere. To generate realistic frequency uncertainties, we searched in the \emph{LEGACY} sample for stars with similar frequency ranges and adopted their observed uncertainties, namely KIC8394589 (Target 1 and 6), KIC12258514 (Target 2), KIC8228742 (Target 3), KIC6508366 (Target 4), and KIC8938364 (Target 5).

\subsection{Results of the hounds}
\label{sec_results_H&H}
To generate the hounds as realistically as possible, we used an advanced modelling strategy which consists first in conducting a Markov Chain Monte-Carlo (MCMC) in a grid, and fitting the classical constraints (effective temperature, metallicity, and luminosity) and the individual frequencies. We used the AIMS software \citep{Rendle2019} to perform the MCMC, and two grids differing by their overshooting value. The global properties of the grids are summarised in Table \ref{tab_properties_grids}, and the physical ingredients of the grid models are the same as for the observed targets in Sec. \ref{sec_behaviour_controlled_environment}, but with one difference that has no influence, partial ionization was not taken into account. With our indicator, we study central stellar regions of main-sequence stars, where accounting for partial ionization or not only has a very small impact \citep{Turcotte1998a,Turcotte1998b}. If radiative accelerations are included or if the target is close to the end of the main sequence, partial ionization is then not negligible \citep{Schlattl2002,Deal2018}. However, inversions are quasi-model-independent, especially they do not rely on the choice of the transport processes. It is therefore not problematic if the reference model does not include the most detailed physics possible for the hare and hounds exercise. Then, we carried out a mean density inversion on the optimal MCMC model to add the inverted mean density to the set of constraints, with a conservative precision of 0.6\%, and we conducted new MCMC fitting the classical constraints, the inverted mean density, and frequency separation ratios ($r_{01}$, $r_{02}$, or both). Finally, we performed local minimizations with a Levenberg-Marquardt algorithm \citep[see e.g.][]{Roweis1996} by leaving the overshooting parameter free. We summarised the constraints in Table \ref{tab_constraints_H&H}, and we included the $r_{01}$ in the constraints of some of the hounds to verify that the correction predicted by the inversion is significantly smaller or negligible, as we expected. This also allows to estimate the impact of the resolution limit. This advanced modelling strategy provided the optimal model of Kepler-93 \citep[which was confirmed by a more elaborated modelling procedure,][]{Betrisey2022}. For stars with physical ingredients that are too different from the physics of the grids, a series of local minimisations are necessary. This was the case for target 5, for which the MCMC was not successful, and for which we conducted local minimisations by starting with guesses of different masses (models \textit{LM, M126, M133, M135,} and \textit{M140}), and with different physical ingredients (models \textit{Ov020/GN93} and \textit{OPLIB/GN93}). In general, if we fit frequency separation ratios, only the $r_{02}$ ratios can be fitted most of the time. Indeed, the fit of both $r_{01}$ and $r_{02}$ ratios tends to be unstable. These ratios are sensitive to different stellar properties, and a minimisation including both of them must deal with too many constraints, preventing a successful convergence onto an optimal solution. An inversion based on the $r_{01}$ ratios is therefore relevant at this stage, because it can provide quasi-model-independent constraints that can highlight favoured and forbidden regions in the parameter space, and provide new insight for the modelling and understanding of the star.

\begin{table}[h!]
\centering
\caption{Global properties of the grids for the MCMC.}
\begin{tabular}{lccc}
\hline 
 & Minimum & Maximum & Step \\ 
\hline \hline 
Mass (M$_\odot$) & 1.20 & 1.50 & 0.02  \\ 
$X_0$ & 0.68 & 0.72 & 0.01 \\ 
$Z_0$ & 0.010 & 0.025 & 0.001 \\ 
Overshooting 1 & \multicolumn{3}{c}{$\rm\alpha_{ov,1}=0.0$} \\ 
Overshooting 2 & \multicolumn{3}{c}{$\rm\alpha_{ov,2}=0.1$} \\
\hline 
\end{tabular} 
\label{tab_properties_grids}
\end{table}

\begin{table}[h!]
\centering
\caption{Constraints for the targets 1, 2, 3, 4, and 6.}
\begin{tabular}{lcc}
\hline 
Label & Constraints & Overshooting \\ 
\hline \hline 
LM & [Fe/H], $T_{\mathrm{eff}}$, $L$, $\bar{\rho}_{Inv}$, $r_{02}$ & free \\ 
LM1 & [Fe/H], $T_{\mathrm{eff}}$, $L$, $\bar{\rho}_{Inv}$, $r_{01}$ & free \\ 
NuOv000 & [Fe/H], $T_{\mathrm{eff}}$, $L$, $\nu_{n,l}$ & $\rm\alpha_{ov}=0.0$ \\ 
NuOv010 & [Fe/H], $T_{\mathrm{eff}}$, $L$, $\nu_{n,l}$ & $\rm\alpha_{ov}=0.1$ \\ 
R01Ov000 & [Fe/H], $T_{\mathrm{eff}}$, $L$, $\bar{\rho}_{Inv}$, $r_{01}$ & $\rm\alpha_{ov}=0.0$ \\ 
R01Ov010 & [Fe/H], $T_{\mathrm{eff}}$, $L$, $\bar{\rho}_{Inv}$, $r_{01}$ & $\rm\alpha_{ov}=0.1$ \\ 
R02Ov000 & [Fe/H], $T_{\mathrm{eff}}$, $L$, $\bar{\rho}_{Inv}$, $r_{02}$ & $\rm\alpha_{ov}=0.0$ \\ 
R02Ov010 & [Fe/H], $T_{\mathrm{eff}}$, $L$, $\bar{\rho}_{Inv}$, $r_{02}$ & $\rm\alpha_{ov}=0.1$ \\ 
RatiosOv000 & [Fe/H], $T_{\mathrm{eff}}$, $L$, $\bar{\rho}_{Inv}$, $r_{01}$, $r_{02}$ & $\rm\alpha_{ov}=0.0$ \\ 
RatiosOv010 & [Fe/H], $T_{\mathrm{eff}}$, $L$, $\bar{\rho}_{Inv}$, $r_{01}$, $r_{02}$ & $\rm\alpha_{ov}=0.1$ \\ 
\hline 
\end{tabular} 
\label{tab_constraints_H&H}
\end{table}

\begin{figure*}[h!]
\begin{subfigure}{.5\textwidth}
  \centering
  \includegraphics[width=.92\linewidth]{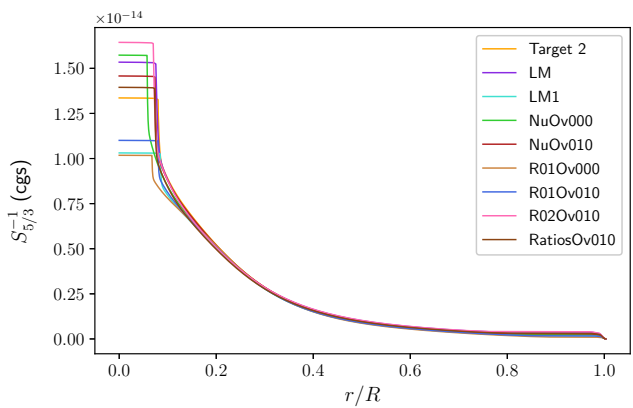}  
  \caption{\centering Entropy profiles of target 2.}
  \label{fig_entropy_target2}
\end{subfigure}
\begin{subfigure}{.5\textwidth}
  \centering
  \includegraphics[width=.95\linewidth]{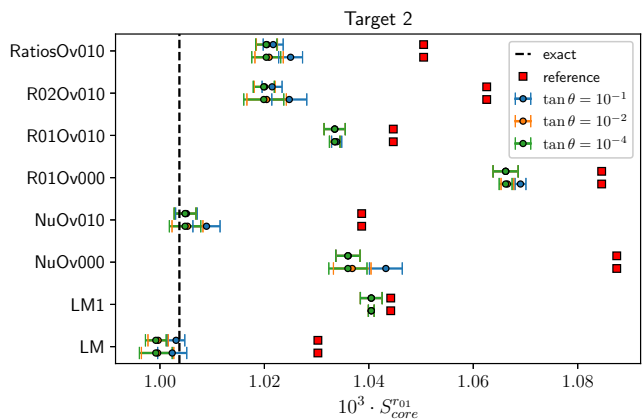}  
  \caption{\centering Hares and hounds results of target 2.}
  \label{fig_target2}
\end{subfigure}
\begin{subfigure}{.5\textwidth}
  \centering
  \includegraphics[width=.92\linewidth]{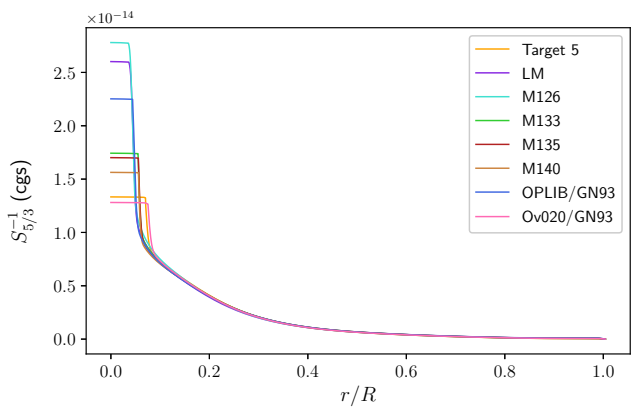}  
  \caption{\centering Entropy profiles of target 5.}
  \label{fig_entropy_target5}
\end{subfigure}
\begin{subfigure}{.5\textwidth}
  \centering
  \includegraphics[width=.95\linewidth]{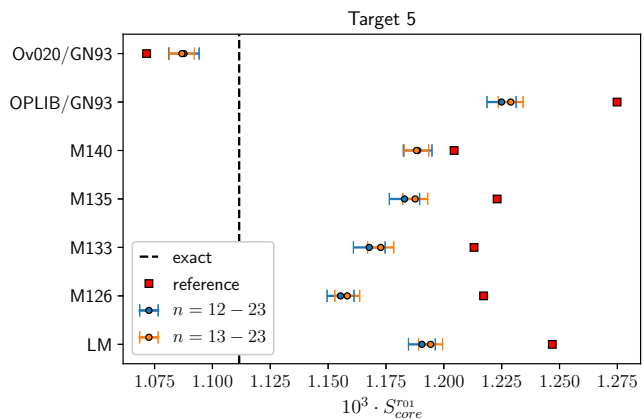}  
  \caption{\centering Hares and hounds results of target 5.}
  \label{fig_target5}
\end{subfigure}
\caption{Entropy profiles and hare and hounds results of targets 2  and 5. The entropy proxy $S_{5/3}$ is in \textit{cgs} units ($\frac{cm^4}{g^{-2/3}s^{-2}}$).}
\label{fig_H&H_interesting_targets}
\end{figure*}

\begin{figure*}[h!]
\begin{subfigure}{.5\textwidth}
  \centering
  \includegraphics[width=.95\linewidth]{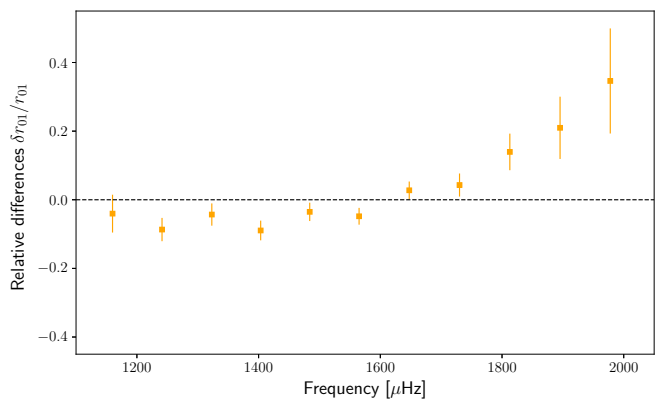}  
  \caption{\centering Model \textit{LM1} of target 2, well fitted $r_{01}$ ratios that are \textit{not} suited for an inversion.}
  \label{fig_ratios_good_fit}
\end{subfigure}
\begin{subfigure}{.5\textwidth}
  \centering
  \includegraphics[width=.95\linewidth]{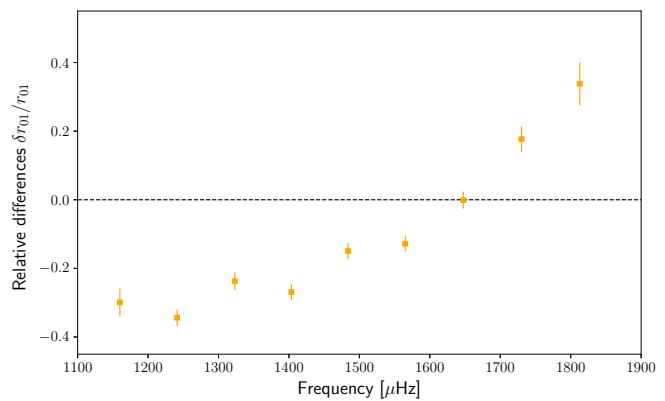}  
  \caption{\centering Model \textit{LM} of target 2, relative $r_{01}$ differences suited for an inversion.}
  \label{fig_ratios_bad_fit}
\end{subfigure}
\caption{Comparison of the relative $r_{01}$ differences for models that can (right) and cannot (left) be used for a meaningful inversion. The relative differences above 0.5 are discarded because they are not appropriate for the linear formalism of the inversion.}
\label{fig_ratios}
\end{figure*}

Two main conditions are required for the inversion to provide a meaningful correction. Here, we already assume that the linear regime assumption is fulfilled, as it is typically the case with the the usual modelling strategies of solar-like stars. First, the indicator must be sensitive to some structural differences, in our case, the differences in the entropy profile in the central stellar regions. Due to our modelling strategy, some of the hounds presented a quasi-identical entropy profile, and were thus identical for our indicator. We therefore discarded the redundant models to avoid biases. In the left panels of Fig. \ref{fig_H&H_interesting_targets}, we illustrate the non-redundant structural differences of the hounds of target 2 and 5. The hare profiles are in orange. The entropy profiles of the hounds differ significantly from each other, the height of the entropy plateau is different, the extent of the convective core is different, and the mixing boundary properties are also different. The corresponding indicator values are displayed in the right panels of Fig. \ref{fig_H&H_interesting_targets} (red squares). We observe that the indicator is indeed sensitive to the structural differences. For example, if the overshooting parameter is changed (e.g. models $NuOv000$ and $NuOv010$), the indicator is significantly affected by the resulting structural differences. Second, if the ratios are well fitted, the relative ratios differences are too small. The inversion combines these relative differences to compute the correction, and if they are too small, the correction is therefore damped. In an ideal case, this behaviour means that the seismic information is well modelled and that additional information cannot be extracted with an inversion. In other words, if the inversion predicts a tiny correction, it confirms that the seismic information is well described by the reference model. However, inversions are numerical methods, and we must ensure that the correction is not driven by numerical noise. In particular, if the ratios are well fitted, the target function may not be well reproduced because of the limited number of modes available. In that case, a non-negligible correction is driven by the bad fit of the target function, and should not be mistaken as the sign that the physics of the reference model is lacking. In Fig. \ref{fig_ratios}, we show examples of relative ratios differences that can and cannot be used for an inversion. We recall that we are conducting linear inversions, and we therefore discarded relative ratios differences that were larger that 0.5 in absolute values, and were therefore not compatible with the linear formalism. Such large relative differences occur for high-order modes and are purely a numerical artefact. For these modes, the frequency separation ratios are close zero and automatically produce a large relative difference. This numerical issue is not limiting, because high-order modes have a very small contribution in the correction predicted by inversion, and can therefore be discarded without loss of information.

The hares fall in two categories of results. For the targets 1, 3, 4, and 6, all the ratios ($r_{01}$ and $r_{02}$) were well fitted, leading to too small relative ratio differences for the inversion. In this situation, the reference value stays within the $1\sigma$-interval of the inverted value, and no new knowledge can be extracted. The most interesting results were obtained for the targets 2 and 5, for which it was not possible to fit simultaneously the $r_{01}$ and $r_{02}$ ratios. In this case, the inversion achieved significant improvements.

\begin{figure*}[h!]
\begin{subfigure}{.5\textwidth}
  \centering
  \includegraphics[width=.87\linewidth]{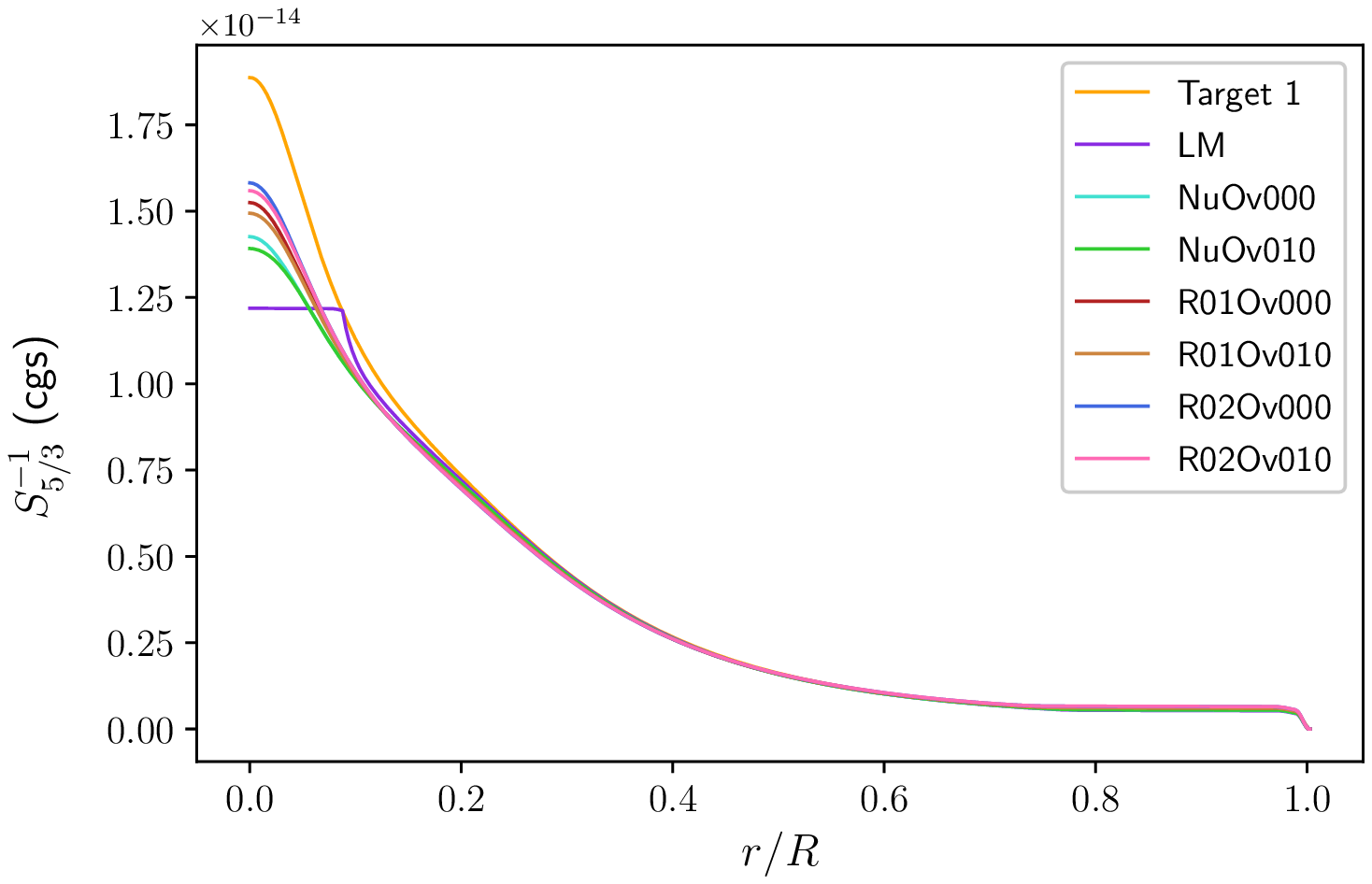}  
  \caption{\centering Entropy profiles of target 1.}
  \label{fig_entropy_target1}
\end{subfigure}
\begin{subfigure}{.5\textwidth}
  \centering
  \includegraphics[width=.9\linewidth]{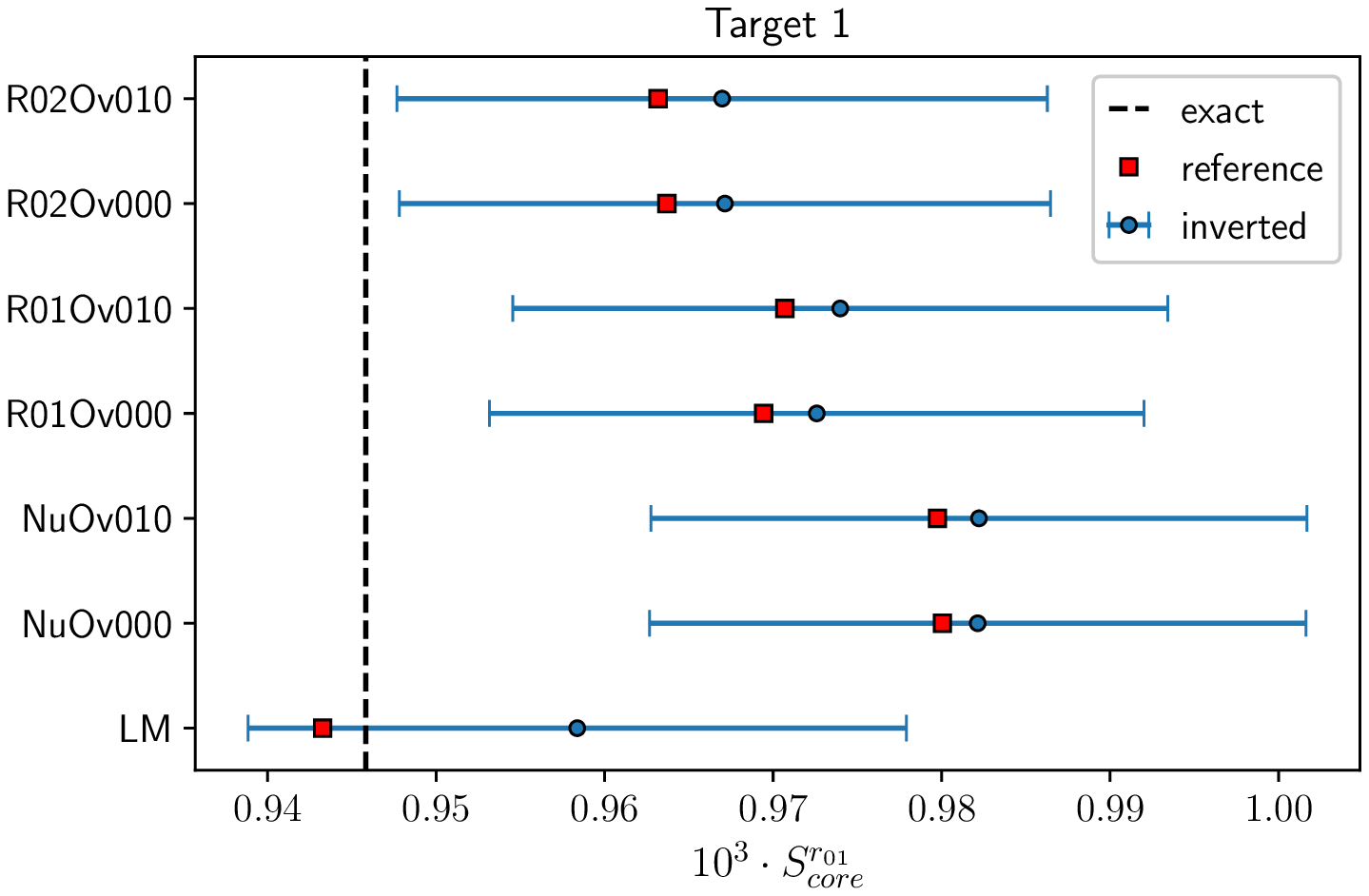}  
  \caption{\centering Hare and hounds results for target 1.}
  \label{fig_target1}
\end{subfigure}
\begin{subfigure}{.5\textwidth}
  \centering
  \includegraphics[width=.87\linewidth]{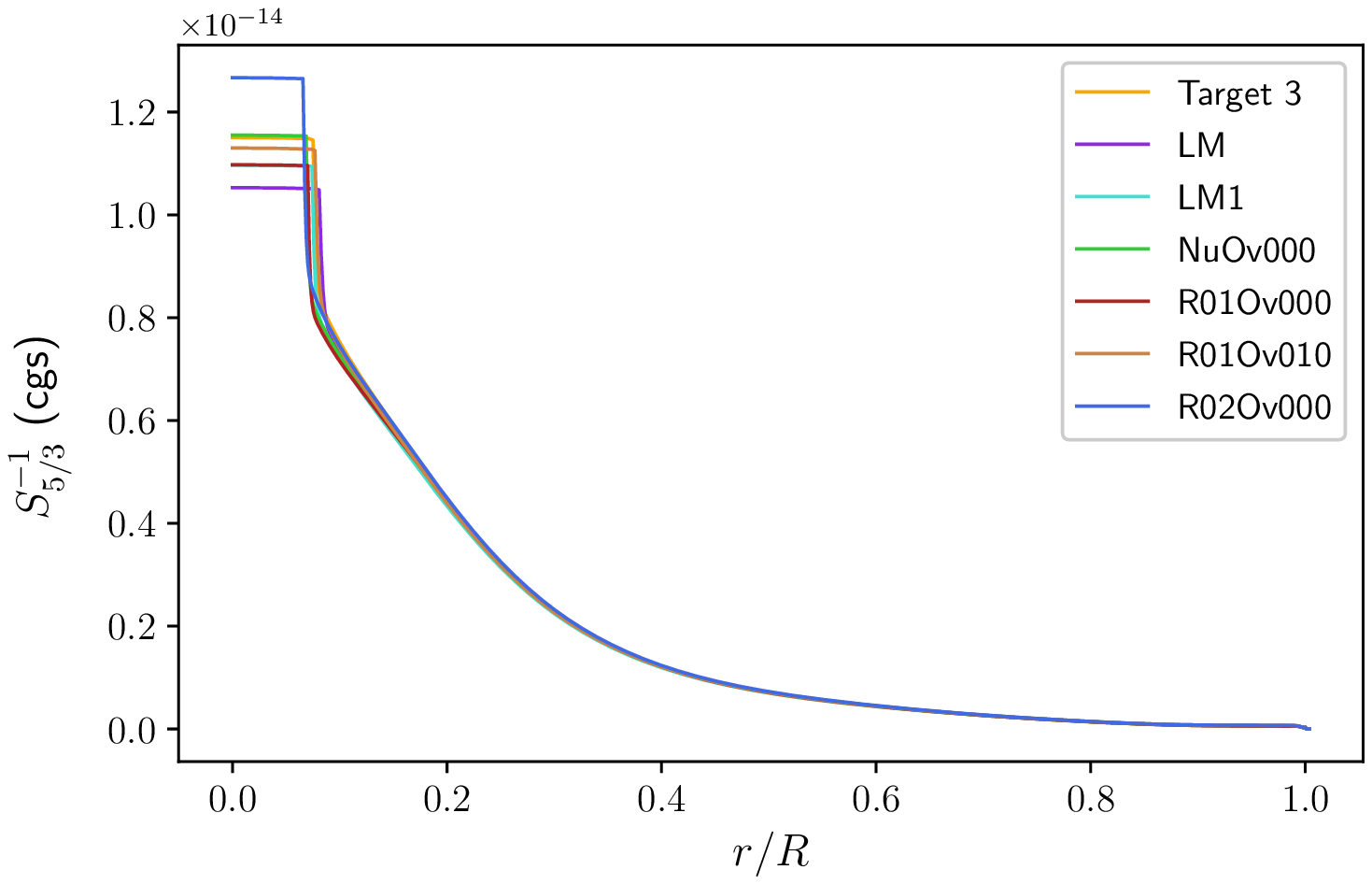}  
  \caption{\centering Entropy profiles of target 3.}
  \label{fig_entropy_target3}
\end{subfigure}
\begin{subfigure}{.5\textwidth}
  \centering
  \includegraphics[width=.9\linewidth]{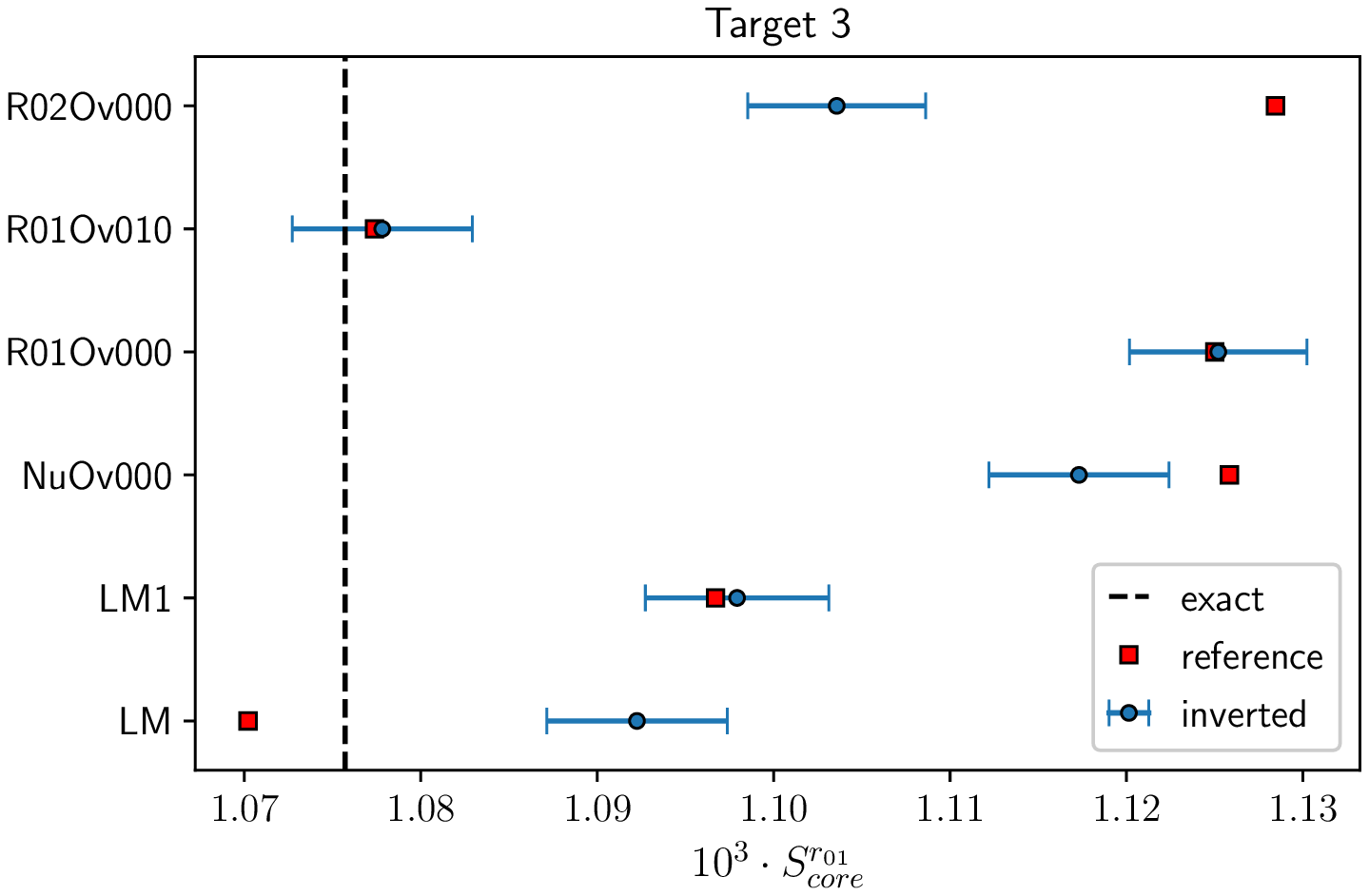}  
  \caption{\centering Hare and hounds results for target 3.}
  \label{fig_target3}
\end{subfigure}
\begin{subfigure}{.5\textwidth}
  \centering
  \includegraphics[width=.87\linewidth]{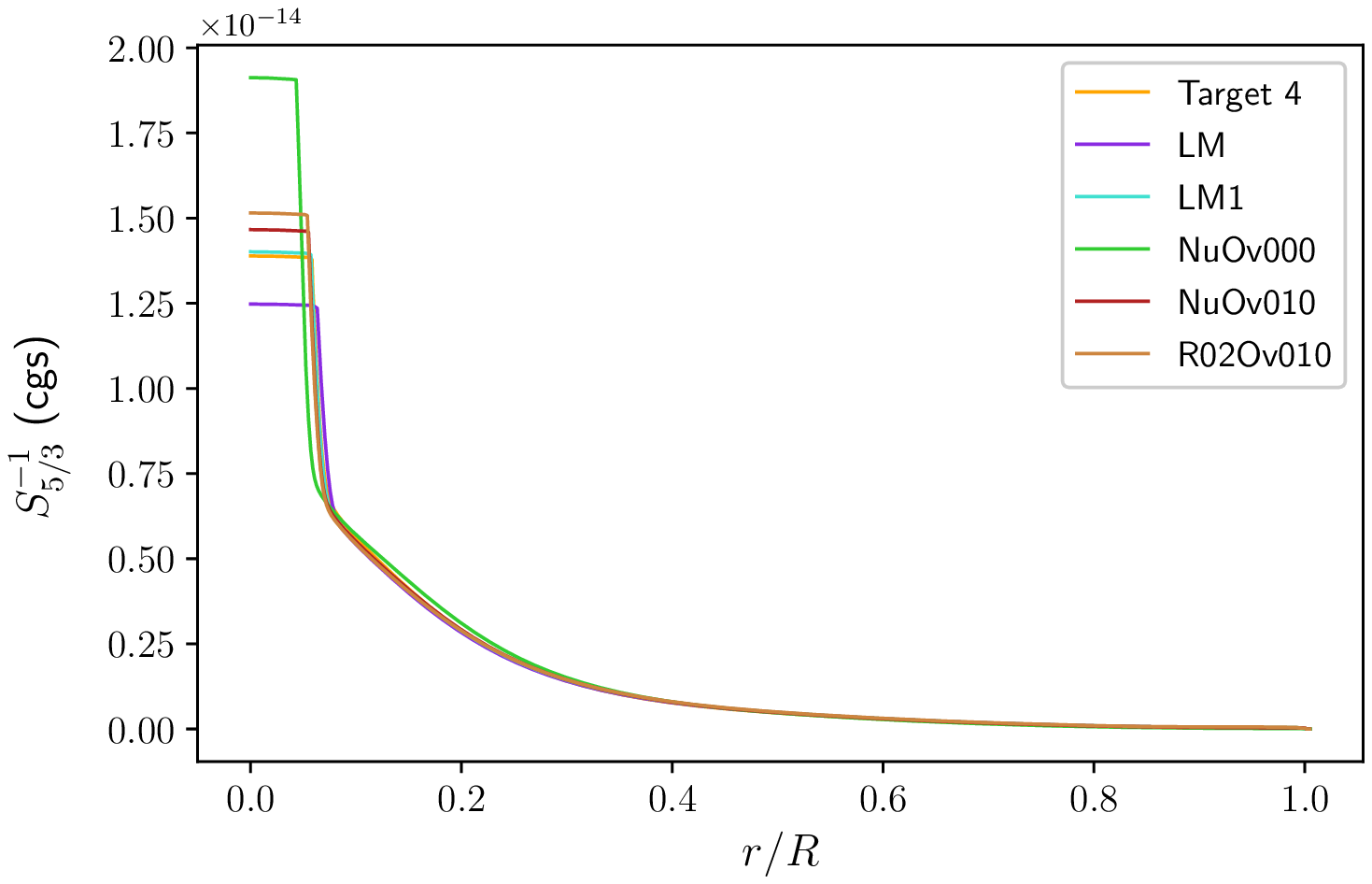}  
  \caption{\centering Entropy profiles of target 4.}
  \label{fig_entropy_target4}
\end{subfigure}
\begin{subfigure}{.5\textwidth}
  \centering
  \includegraphics[width=.9\linewidth]{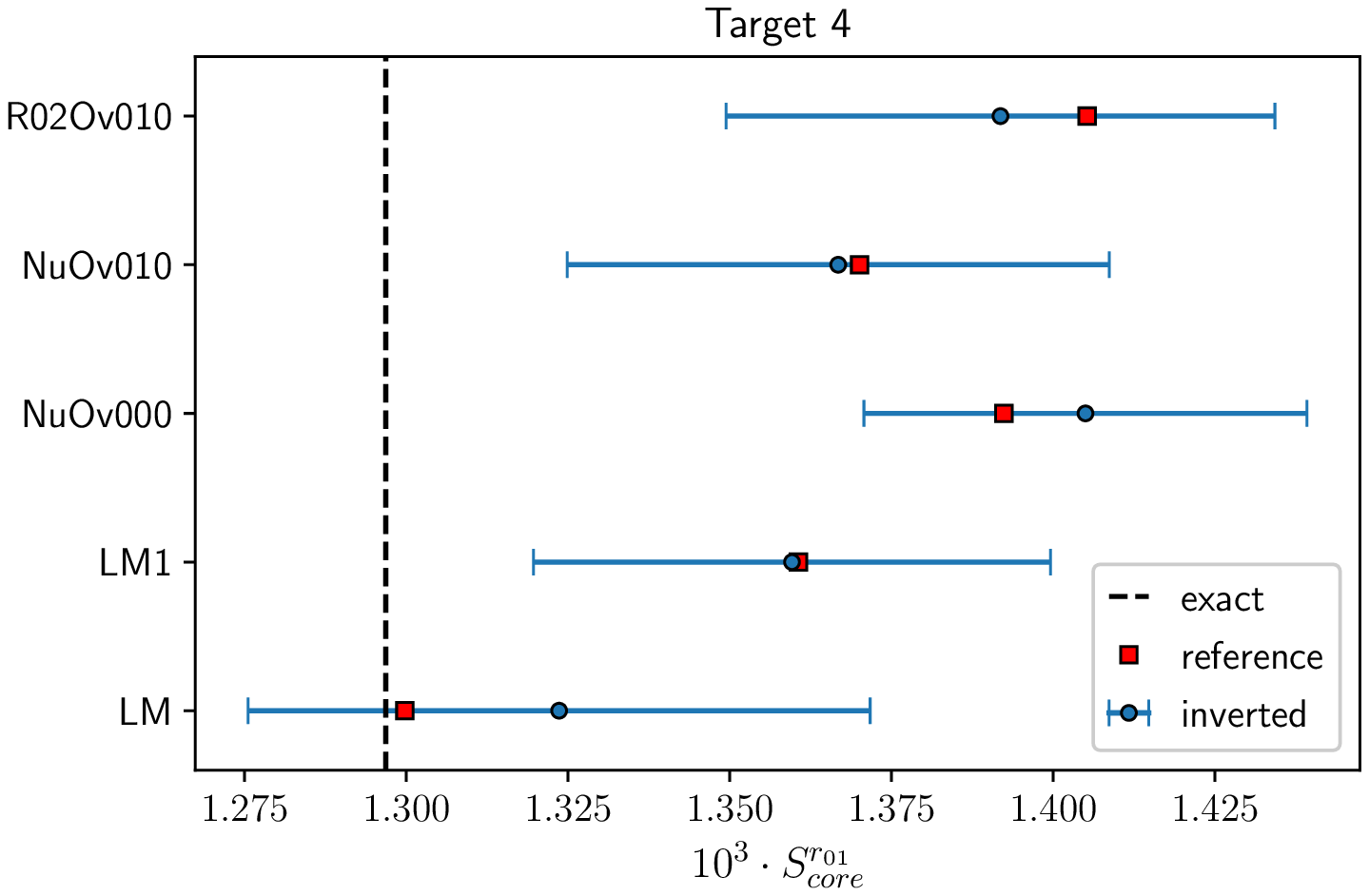}  
  \caption{\centering Hare and hounds results for target 4.}
  \label{fig_target4}
\end{subfigure}
\begin{subfigure}{.5\textwidth}
  \centering
  \includegraphics[width=.87\linewidth]{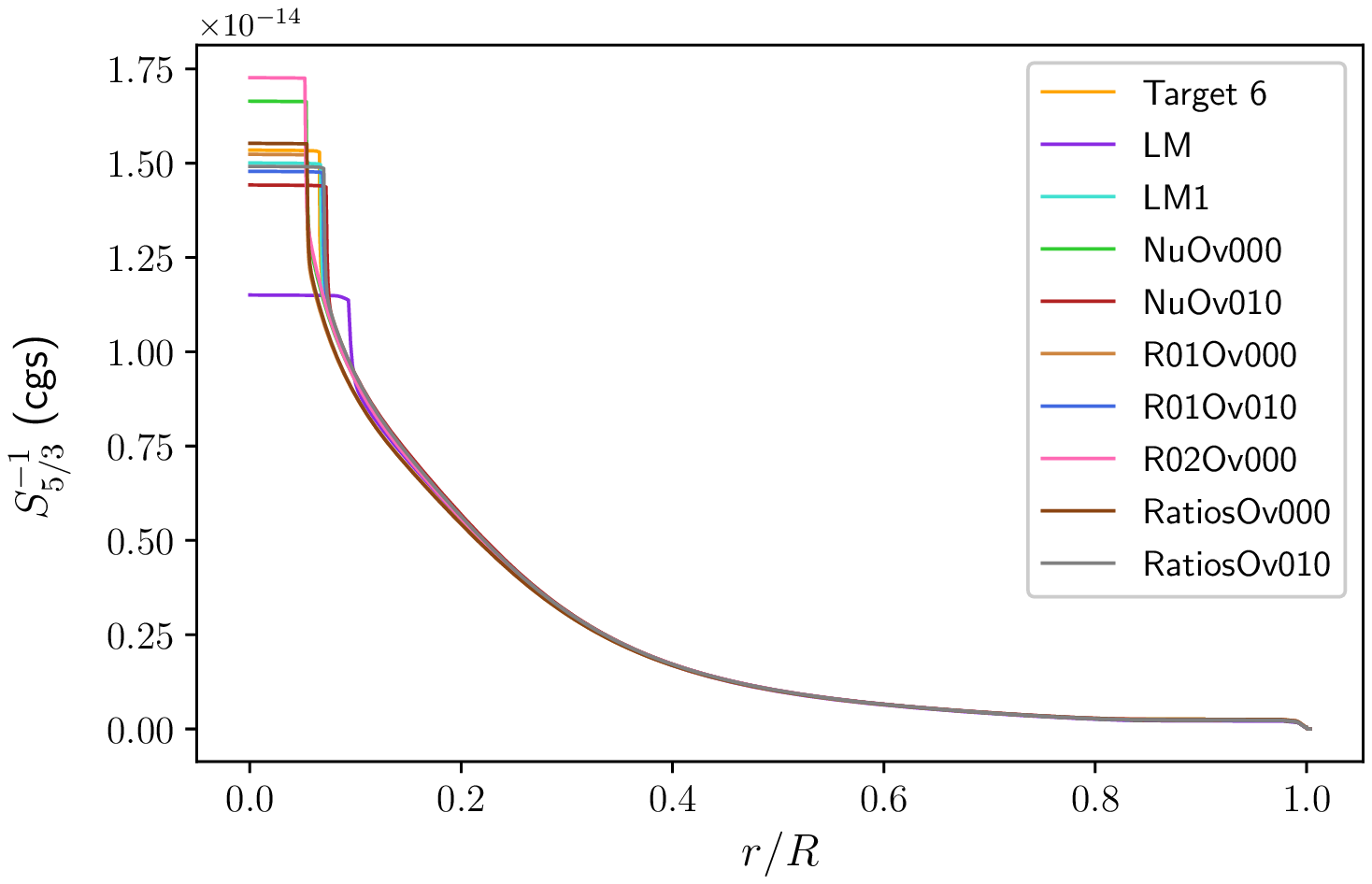}  
  \caption{\centering Entropy profiles of target 6.}
  \label{fig_entropy_target6}
\end{subfigure}
\begin{subfigure}{.5\textwidth}
  \centering
  \includegraphics[width=.9\linewidth]{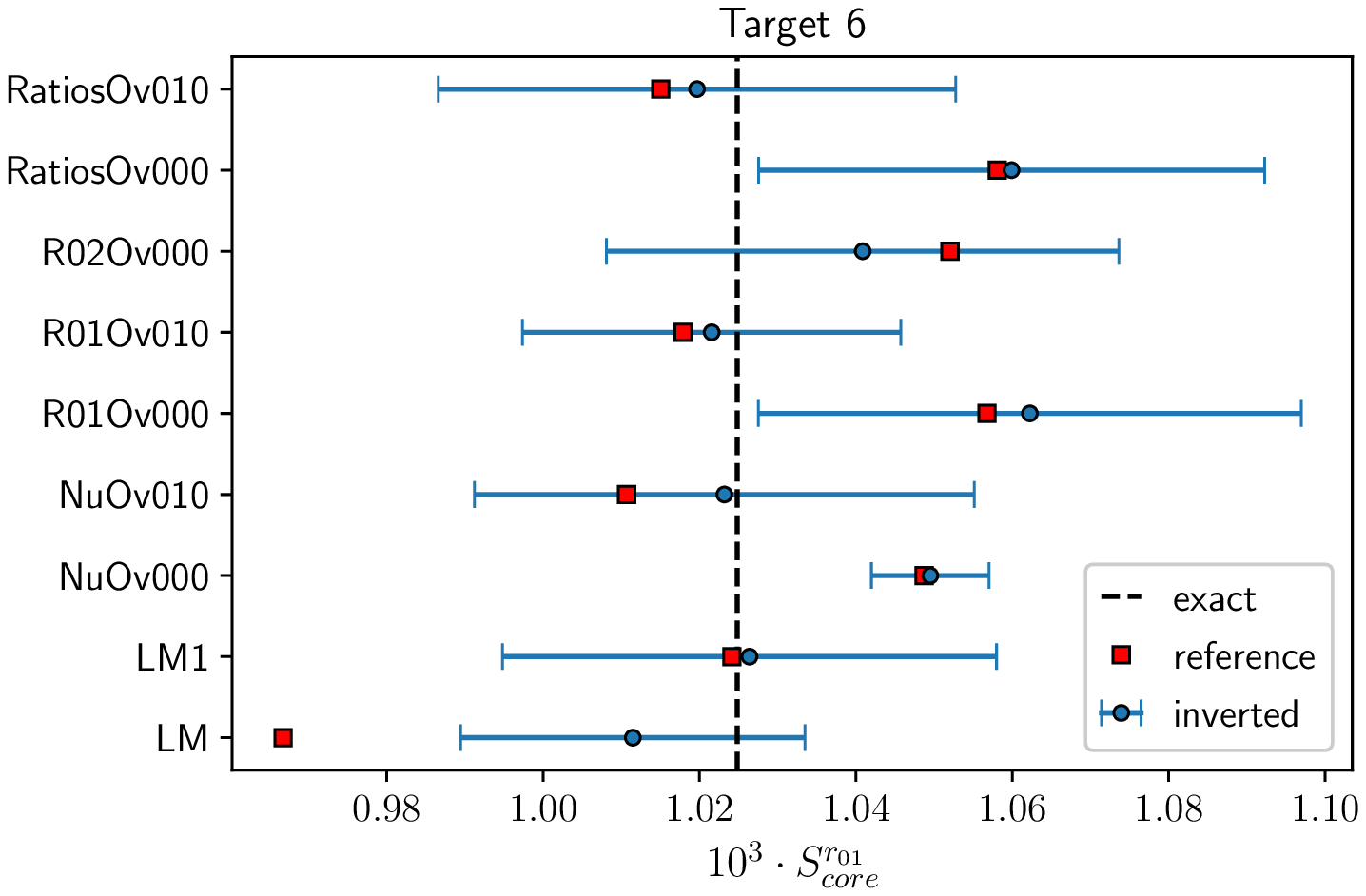}  
  \caption{\centering Hare and hounds results for target 6.}
  \label{fig_target6}
\end{subfigure}
\caption{Entropy profiles and hare and hounds results of targets 1, 3, 4, and 6. The entropy proxy $S_{5/3}$ is in \textit{cgs} units ($\frac{cm^4}{g^{-2/3}s^{-2}}$).}
\label{fig_H&H_ratios_well_fitted}
\end{figure*}

The entropy profiles and results for the first category of targets are shown in Fig. \ref{fig_H&H_ratios_well_fitted}. Target 3 shows the reason for which we stressed out that an inversion should not be carried out on well fitted ratios. Theoretically, we would expect a negligible correction in such conditions, as for targets 1 and 4. However, some of the hounds of target 3 exhibit non-negligible corrections, which are a combination of physical structural differences and the resolution limit of the inversion. We observe that the inversion corrects in the right direction, but its magnitude is difficult to trust, because it is probably largely dominated by the quality of the fit of the target function. The behaviour of the inversions for target 6 is similar to the results for targets 1 and 4, with one exception. The ratios were not well fitted by the \textit{LM} model, and the entropy profiles were significantly different (Fig. \ref{fig_entropy_target6}). The conditions for a meaningful inversion were therefore met, and the inversion behaves accordingly with our expectations, as illustrated in Fig. \ref{fig_target6}.

The results for the targets 2 and 5 are displayed in Figs. \ref{fig_target2} and \ref{fig_target5}, respectively. Because the relative ratio differences were significant enough to carry out an inversion, we conducted more intensive tests. For target 2, we tested the impact of the trade-off parameter $\theta$, and our choice of $\theta=10^{-4}$ appears as a good compromise. Indeed, only a significantly higher value can slightly affect the inversion result, and in that case, the target function was less well reproduced because it corresponds to assuming high observational uncertainties. We also tested two uncertainty sets, where we considered the standard observational uncertainties and checked that equal-weighted uncertainties produce similar results. In this fashion, we could discard the possibility of the presence of highly non-linear modes. For target 5, we tested two mode sets by removing the lowest order mode. For both sets, the inversion significantly improves the reference values, which shows that it is not necessary to use too low order modes that could be difficult to observe. Moreover, model \textit{Ov020/GN93} strengthens the consistency of the inversion, because it lies at the opposite side of the exact value with respect to the other models, and therefore corrects in the opposite direction. We point out that even if we do not have the same convection theory in the hounds of target 5, which strongly impacts the stellar structure, the inversion is still able to correct significantly and in the right direction.

In Table \ref{tab_results_targets_2_5}, we tried to estimate the improvement achieved by the inversion. We computed the average value\footnote{For target 2, we excluded the model \textit{LM1} because it was a fit of the $r_{01}$ ratios, which made the inversion meaningless. We still kept the other models including the $r_{01}$ in the constraints because they produced relative ratio differences large enough to carry out an inversion.} of the indicator, and defined an accuracy measure as the relative difference of the reference/inverted value with respect to the exact value. For target 2, the accuracy is improved from 5.3\% to 2.2\%, and for target 5, from 8.6\% to 5.4\%. This result should be put in perspective with the precision achieved by the inversion. From a pure statistical point of view, the precision is of the order $\sim 0.1\%$ to $0.2\%$, which is negligible in comparison to the systematic uncertainty due to the choice of the physical ingredients, and the resolution limit of the inversion. In practice, a modelling strategy similar to that of \citet{Betrisey2022} should be considered. An extensive set of models with various physical ingredients should be constructed, and the precision is measured with the standard deviation of the inversion results. With this approach, the precision is significantly lower than the statistical precision, but it is generally smaller or equal to the precision of the reference models. Hence, the inversion can shift and in some cases shrink the reference range towards the exact value.

\begin{table}
\centering
\caption{Results for the targets 2 and 5.}
\begin{tabular}{lccc}
\hline 
 & $S_{core}^{r_{01}}$ &  Accuracy & Precision \\ 
\hline \hline 
\textit{Target 2} &  & \\ 
Reference & $1.057\cdot 10^{-3}$  & 5.3\% & 2.1\% \\ 
Inverted & $1.026\cdot 10^{-3}$ & 2.2\% & 2.2\%\\ 
Exact & $1.004\cdot 10^{-3}$ & &  \\ 
\hline \hline
\textit{Target 5} & &  \\ 
Reference & $1.207\cdot 10^{-3}$ & 8.6\% & 5.3\% \\  
Inverted & $1.171\cdot 10^{-3}$ & 5.4\% & 3.6\% \\ 
Exact & $1.112\cdot 10^{-3}$ & &  \\ 
\hline 
\end{tabular} 
\label{tab_results_targets_2_5}
\end{table}


\section{Conclusions}
\label{sec_conclusions}
In this work, we adapted the SOLA method for frequency separation ratios in Sec. \ref{sec_theoretical_considerations}, and introduced a new indicator probing the properties of stellar convective cores in Sec. \ref{sec_definition_indicator}. Thanks to this approach, the inversion is not affected by surface effects. In Sec. \ref{sec_behaviour_controlled_environment} we verified our technique in a controlled environment, where the stellar mass and radius are identical between the reference and observed models. Then, we conducted an intensive hare and hounds exercise in Sec. \ref{sec_H&H}.

Our inversion technique showed promising results to probe the convective core of main-sequence stars, and is especially suited for F-type stars. Indeed, they have an adequate mass span for the indicator, and the inversion is not sensitive to the surface effects that are difficult to treat for this type of stars. Because the entropy differences are localised in the very central stellar regions that are difficult to probe even with the ratios, we found that a modelling strategy similar to that of \citet{Betrisey2022} should be considered in a practical application of the method. An extensive set of models should be constructed, and inferences about the properties of the core can be deducted from the global behaviour of the inversion on the models of the set. The ratios are influenced by the overshooting, the radiative opacities, and the chemical composition. It is therefore not possible to constrain directly one of these quantities, but favoured and forbidden regions can be highlighted in the parameter space. We stress out that if the ratios are well reproduced, an inversion is meaningless due to the resolution limit of the method.

In a broader modelling context, the inversion provides quasi-model-independent constraints on core properties that can help to point out limitations of the reference models, and for example be the sign of an incorrect or missing physical process. The inversion can also help to exclude some of the reference models, thus improving the accuracy and precision of stellar fundamental parameters such as mass, radius, and age, a crucial objective for the PLATO mission.


\section*{Acknowledgements}
J.B and G.B. acknowledge fundings from the SNF AMBIZIONE grant No 185805 (Seismic inversions and modelling of transport processes in stars).


\bibliography{bibliography.bib}

\appendix

\section{The $r_{10}$ and $r_{02}$ kernels}
\label{sec_appendix_kernel_r10_r02}
For the $r_{02}$ ratios, we have:
\begin{align}
\frac{\delta r_{02}}{r_{02}}(n) &= \frac{\delta d_{02}(n)}{d_{02}(n)} - \frac{\delta\Delta_1(n)}{\Delta_1(n)}, \\
                                                        &= \int_0^R\left(K_{a,b}^{r_{02}}(n)\frac{\delta a}{a} + K_{b,a}^{r_{02}}(n)\frac{\delta b}{b}\right)dr
                                                               + \mathcal{O}(\delta^2),
\end{align}
where:
\begin{align}
K_{a,b}^{r_{02}}(n) &= \frac{\nu_{n,0}K_{a,b}^{n,0}-\nu_{n-1,2}K_{a,b}^{n-1,2}}{\nu_{n,0}-\nu_{n-1,2}}
                                           - \frac{\nu_{n,1}K_{a,b}^{n,1}-\nu_{n-1,1}K_{a,b}^{n-1,1}}{\nu_{n,1}-\nu_{n-1,1}}.
\end{align}

The kernel $K_{b,a}^{r_{02}}(n)$ is found by switching $a$ and $b$ in the previous expression.

For the $r_{10}$ ratios, we have:
\begin{align}
\frac{\delta r_{10}}{r_{10}}(n) &= \frac{\delta d_{10}(n)}{d_{10}(n)} - \frac{\delta\Delta_0(n+1)}{\Delta_0(n+1)}, \\
                                                        &= \int_0^R\left(K_{a,b}^{r_{10}}(n)\frac{\delta a}{a} + K_{b,a}^{r_{10}}(n)\frac{\delta b}{b}\right)dr
                                                               + \mathcal{O}(\delta^2),
\end{align}
where:
\begin{small}
\begin{align}
K_{a,b}^{r_{10}}(n) = &\frac{\nu_{n-1,1}K_{a,b}^{n-1,1}-4\nu_{n,0}K_{a,b}^{n,0}+6\nu_{n,1}K_{a,b}^{n,1}
                                                  -4\nu_{n+1,0}K_{a,b}^{n+1,0}+\nu_{n+1,1}K_{a,b}^{n+1,1}}{\nu_{n-1,1}-4\nu_{n,0}+6\nu_{n,1}
                                                  -4\nu_{n+1,0}+\nu_{n+1,1}} \nonumber\\
                                        &   - \frac{\nu_{n+1,0}K_{a,b}^{n+1,0}-\nu_{n,0}K_{a,b}^{n,0}}{\nu_{n+1,0}-\nu_{n,0}}.
\end{align}
\end{small}

Again, the kernel $K_{b,a}^{r_{10}}$ is found by switching $a$ and $b$ in the previous expression.

\section{Correlation matrix for the ratios}
\label{sec_appendix_correlation_matrix_ratios}
Let $X_i$ be $N$ random variables, and $X_i^0$ be their respective mean. We consider the random variable $Z$ which is the general form of a frequency ratio:
\begin{equation}
Z = \frac{\sum_{i=1}^N a_i X_i}{\sum_{i=1}^N b_i X_i}.
\end{equation}

In order to derive the error propagation, we use the first order Taylor expansion around the mean,
\begin{tiny}
\begin{align}
Z(X_1,...,X_n) &\simeq Z(X_1^0,...,X_n^0) + \sum_{i=1}^N\frac{\partial Z}{\partial X_i}(X_1^0,...,X_n^0)(X_i-X_i^0), \\
                           &= Z(X_1^0,...,X_n^0) + \sum_{i=1}^N\frac{\left(a_i -Z(X_1^0,...,X_n^0) b_i\right)\left(X_i-X_i^0\right)}{\sum_k b_k X_k^0}.
\end{align}
\end{tiny}

In the following, we introduce the short notation $Z^0 \equiv Z(X_1^0,...,X_n^0)$. At first order, the correlation matrix $\mathrm{Cov}(Z,\hat{Z})$ is equal to
\begin{tiny}
\begin{align}
& \mathrm{Cov}\left(Z^0 + \sum_{i=1}^N\frac{(a_i -Z^0 b_i)(X_i-X_i^0)}{\sum_k b_k X_k^0},\ \hat{Z}^0 + \sum_{j=1}^N\frac{(\hat{a}_j -\hat{Z}^0 \hat{b}_j)(X_j-X_j^0)}{\sum_k \hat{b}_k X_k^0}\right), \\
&=\mathrm{Cov}\left(\sum_{i=1}^N\frac{a_i -Z^0 b_i}{\sum_k b_k X_k^0}X_i,\ \sum_{j=1}^N\frac{\hat{a}_j -\hat{Z}^0 \hat{b}_j}{\sum_k \hat{b}_k X_k^0}X_j\right), \\
&= Z^0\hat{Z}^0\sum_{i=1}^N\sum_{j=1}^N\left(\frac{a_i}{\sum_k a_k X_k^0}-\frac{b_i}{\sum_k b_k X_k^0}\right)\left(\frac{\hat{a}_j}{\sum_k \hat{a}_k X_k^0}-\frac{\hat{b}_j}{\sum_k \hat{b}_k X_k^0}\right)\mathrm{Cov}(X_i,X_j),
\end{align}
\end{tiny} \unskip
where we used the linearity properties of the covariance matrix, and that the covariance of a random variable with a constant is null.

For a set of $N_f$ frequencies $\nu_i$, whose uncertainties are described by the covariance matrix $E_{ij}$, and $i$ being a short notation for the identification pair $(n,l)$, a frequency ratio can be expressed as:
\begin{equation}
r = \frac{\sum_{i=1}^{N_f} a_i \nu_i}{\sum_{i=1}^{N_f} b_i \nu_i}.
\end{equation}
For example, for the $r_{02}(n)$ ratio, the coefficients are:
\begin{tiny}
\begin{align}
\vec{a} &= [0_{(n_{min},0)},...,0_{(n-1,0)},1_{(n,0)},0_{(n+1,0)},..., \nonumber\\
               &\quad\enspace 0_{(n-2,2)},-1_{(n-1,2)},0_{(n,2)},...,0_{(n_{max},l_{max})}] \\
\vec{b} &= [0_{(n_{min},0)},...,0_{(n-2,1)},-1_{(n-1,1)},1_{(n,1)},0_{(n+1,1)},...,0_{(n_{max},l_{max})}],
\end{align}
\end{tiny} \unskip
where the index indicates the identification pair of the corresponding frequency.

The covariance between two ratios is given by
\begin{tiny}
\begin{equation}
\mathrm{Cov}(r,\hat{r}) = r\hat{r}\sum_{i=1}^{N_f}\sum_{j=1}^{N_f}\left(\frac{a_i}{\sum_k a_k \nu_k}-\frac{b_i}{\sum_k b_k \nu_k}\right)\left(\frac{\hat{a}_j}{\sum_k \hat{a}_k \nu_k}-\frac{\hat{b}_j}{\sum_k \hat{b}_k \nu_k}\right)E_{ij}.
\end{equation}
\end{tiny}

If the frequencies are uncorrelated, which is the typical assumption, one has $E_{ij}=\sigma_i^2\delta_{ij}$, where $\sigma_i$ is the uncertainty of the frequency $\nu_i$. In this case, the covariance simplifies to
\begin{tiny}
\begin{equation}
\mathrm{Cov}(r,\hat{r}) = r\hat{r}\sum_{i=1}^{N_f}\left(\frac{a_i}{\sum_k a_k \nu_k}-\frac{b_i}{\sum_k b_k \nu_k}\right)\left(\frac{\hat{a}_i}{\sum_k \hat{a}_k \nu_k}-\frac{\hat{b}_i}{\sum_k \hat{b}_k \nu_k}\right)\sigma_i^2,
\label{eq_cov_ratios}
\end{equation}
\end{tiny} \unskip
and the $1\sigma$ error bar on $r$ is
\begin{equation}
\sigma_r^2 = r^2\sum_{i=1}^{N_f}\left(\frac{a_i}{\sum_k a_k \nu_k}-\frac{b_i}{\sum_k b_k \nu_k}\right)^2\sigma_i^2.
\end{equation}

In the SOLA method, we use relative differences of ratios,
\begin{equation}
\frac{\delta r}{r} = \frac{r_{obs}-r_{mod}}{r_{mod}},
\end{equation}
where $obs$ stands for observed and $mod$ for modelled. Eq. \eqref{eq_cov_ratios} therefore becomes
\begin{tiny}
\begin{equation}
\mathrm{Cov}\left(\frac{\delta r}{r},\hat{\frac{\delta r}{r}}\right) = \frac{r_{obs}}{r_{mod}}\frac{\hat{r}_{obs}}{\hat{r}_{mod}}\sum_{i=1}^{N_f}\left(\frac{a_i}{\sum_k a_k \nu_k}-\frac{b_i}{\sum_k b_k \nu_k}\right)\left(\frac{\hat{a}_i}{\sum_k \hat{a}_k \nu_k}-\frac{\hat{b}_i}{\sum_k \hat{b}_k \nu_k}\right)\sigma_i^2.
\label{eq_entries_cov_matrix_ratios}
\end{equation}
\end{tiny} \unskip
The error term in the SOLA cost function is then
\begin{equation}
\tan\theta\sum_{p=1}^{N_r}\sum_{q=1}^{N_r} \frac{c_p c_q \vec{\Sigma}_{pq}}{<\vec{\Sigma}^2>},
\end{equation}
where $\vec{\Sigma}_{pq}=\mathrm{Cov}\left((\frac{\delta r}{r})_p,(\frac{\delta r}{r})_q\right)$, $<\vec{\Sigma}^2> = \frac{1}{N_r^2}||\vec{\Sigma}||_\mathrm{F}^2$, $||\vec{\Sigma}||_\mathrm{F}^2$ is the square of the Frobenius norm of the covariance matrix of the ratios, and $N_r$ is the number of observed ratios.

\end{document}